\documentclass{article}
\usepackage{graphics}
\def\no{\noindent}

\def\bc{\begin{center}}
\def\ec{\end{center}}

\def\L{\Lambda}
\def\beq{\begin{equation}}
\def\eeq{\end{equation}}
\def\no{\noindent}
\def\z{\zeta}

\def\s{\sigma}
\def\om{\omega}
\def\vphi{\varphi}

\def\bpsi{{\bar\psi}}

\begin{document}
\title{Phase Transitions of a Bose Gas in an Optical Lattice}

\maketitle

\author{K. Ziegler, Institut f\"ur Physik, Universit\"at Augsburg, Germany}

\no
Abstract:

\no
A grand-canonical system of interacting bosons is considered to study
phase transitions of ultracold atoms in an optical lattice.
The phase diagram is discussed in terms of a matrix-like order parameter,
representing a symmetric phase (Mott insulator) and a symmetry-broken phase
(Bose-Einstein condensate). An exact solution was found by introducing $N$ 
colors and taking the limit $N\to\infty$.   

\section{\bf Introduction }

In this paper an interacting Bose gas in an optical lattice is studied.
It is based on some previously published ideas 
for the hard-core Bose gas and its generalization to soluble models. 
The application to Bose-Einstein condensation, depletion of the condensate
and the Mott insulating state in these systems is discussed.

A real Bose gas, as has been studied in the case of trapped cold alkali
atoms \cite{anderson}, is always subject to interparticle interaction. 
In a good
approximation this interaction is a hard core, characterized by a scattering
length $a$. Depending on the type of atoms and external conditions
(e.g. absence or presence of an external magnetic field), the scattering
length can vary over a wide range. Typical values of $a$ observed in 
experiments are $a\approx 5$nm in $^{87}$Rb \cite{newbury95} and up to
$a\approx 500$nm in a gas of $^{85}$Rb atoms near the Feshbach resonance 
\cite{courteille}. Negative scattering lengths are also known but will not 
be considered here. The effect of the hard-core interaction can be
characterized by the interaction parameter
\[
I={{\rm scattering\  length}\over{\rm typical\  interparticle\  distance}}.
\]
If the typical interparticle distance in a Bose gas is large (e.g., 
$\approx 200$nm in a very dilute gas that undergoes a Bose-Einstein
condensation \cite{anderson}), $I$ is small ($I\approx 0.01$), and the 
interaction can be
treated as a weak perturbation. This is the regime where the Gross-Pitaevskii
physics can be used \cite{ginzburg,dalfovo}.

If the Bose gas is transferred to an optical lattice the characteristic
length scale is given by the distance of the potential minima, the 
lattice constant $a_l$. Typical experimental values are $a_l\approx 
250...500$nm. Then the interaction is $I\approx 0.01...1$.


\no 
>From the theoretical standpoint the simplest case is the
non-interacting Bose gas ($I=0$), where no qualitative effect of the
lattice on the Bose-Einstein condensation is expected.
Indeed, Bose-Einstein condensation in a non-interacting Bose gas is not a 
phase transition, characterized by spontaneous symmetry breaking, but a
macroscopic occupation of the groundstate.
A grand-canonical ensemble of non-interacting bosons at inverse temperature
$\beta$, with quantum numbers $l=0,1,...$ and 
energy $\epsilon_l\ge0$ is given by the partition function
\beq
Z_0=Tre^{-\beta {\hat H}}
=\prod_{l\ge0}\sum_{ n_l\ge0}(\zeta e^{-\epsilon_l})^{\beta n_l}.
\label{partition}
\eeq
The fugacity
$\zeta$ is chosen such that $\epsilon_0=0$ and $\epsilon_l>0$ for $l>0$.
It is assumed that the groundstate is unique.
Then $\zeta>1$ is not allowed here because this would lead to a divergent
series in Eq. (\ref{partition}) for $l=0$.
The average number of particles with quantum number $l$ is
\beq
N_l=\langle n_l\rangle=-{1\over\beta}{\partial\over\partial\epsilon_l}\log Z
={1\over e^{\beta\epsilon_l}\zeta^{-\beta}-1}.
\label{number}
\eeq
This quantity decreases with increasing $\epsilon_l$ and has its maximum at 
$l=0$. The probablity of finding a particle in the groundstate is
\beq
{N_0\over N}=1-{1\over N}\sum_{l\ge 1} N_l.
\label{occ}
\eeq
The Bose-Einstein condensate (BEC) is the macroscopically occupied
groundstate, i.e., $\lim_{N\to\infty}N_0/N>0$. 
According to the result (\ref{number}) this can only happen if
$\zeta$ goes to 1 when $N$ goes to infinity. Using the density of states
$\rho(\epsilon)$ at energy $\epsilon$
and keeping the ratio $N/V$ ($V$: volume of the Bose gas) fixed when 
$N\to\infty$, the second term in Eq. (\ref{occ}) becomes the integral
\[
{V\over N\beta}
\int_0^\infty{\rho(x/\beta)\over e^x-1}dx.
\]
There exists a BEC if the integral is smaller than 1. The latter is the case
if (i) the density of states $\rho(\epsilon)$ cuts-off the pole of
the integrand (this is usually the case for 
three-dimensional lattices) and if (ii) $V/(N\beta)$ is sufficiently small.
Thus the presence of a lattice does not prohibit the formation of a
BEC in a non-interacting Bose gas at sufficiently low temperatures.
More interesting physics is expected for $I\approx1$ when the lattice
constant is of the same order as the scattering length because the
interaction can block the tunneling processes and lead to an
insulating state without phase coherence and with a locally fixed
number of bosons, the Mott insulator.
 
\section{\bf Random-Walk Representation}

\no
In order to discuss the interaction between bosons in an optical lattice it is
convenient to derive a random-walk representation for the lattice Bose gas.
This is related to the idea that the average over particle fluctuations in an
equilibrium system is equivalent to the average of random walks.
For this purpose one can start from non-interacting bosons, then develop the
weakly (Bose-Hubbard-like) and finally the strongly (hard-core) interacting 
gas.
\subsection{Non-interacting Bose Gas}

Evaluation of the geometric series in the partition function of Eq. 
(\ref{partition}) gives
\[
Z_0
=\prod_l{1\over 1-\zeta^\beta e^{-\beta\epsilon_l}}.
\]
Assuming that $\beta$ is an integer, the identity
$$
1-\zeta^\beta e^{-\beta\epsilon_l}=\det({\bf 1}-\zeta w_l)
$$
with the $\beta\times\beta$ matrix
\beq
w_l=\pmatrix{
0 & u_l & 0 & \ldots & 0 \cr
0 & 0  & u_l & \ddots & \vdots \cr
\vdots & \ddots & \ddots & \ddots & 0 \cr
0      & \ldots & 0 & 0 & u_l \cr
u_l  & 0 & \ldots & 0 & 0 \cr
}\ \ \ \ (u_l=e^{-\epsilon_l})
\label{matrix}
\eeq
and the diagonal $\beta{\cal N}\times\beta{\cal N}$ matrix 
${\bf w}=diag(w_1,w_2,...,w_{\cal N})$ can be used to write the partition 
function as 
\beq
Z_0={1\over\det({\bf 1}-\zeta {\bf w})}.
\label{fint}
\eeq
The matrix elements in the last equation are defined with respect to $x=(l,t)$
($t=1,2,...,\beta$, $l=0,1,...,{\cal N}$).
The diagonal matrix ${\bf w}$ can be replaced by any other matrix to 
describe a non-interacting Bose gas. For instance, tunneling between
sites $r,r'$ of the three-dimensional optical lattice gives with 
$x=(r,t)$ the matrix elements
\beq
w_{x,x'}=\cases{
J\delta_{t',t+1}/6 & for $r,r'$ nearest neighbors \cr
(1-J)\delta_{t',t+1} & for $r'=r$ \cr
0 & otherwise \cr
}.
\label{tunnel}
\eeq
This choice implies that a particle either stays in a potential well of
the optical lattice at $r$
with probability $1-J$ ($0\le J\le1$) or tunnels from a well at $r$ to one 
of the six neighboring wells with probability $J/6$. 
An expansion of $Z_0$ in powers of $\zeta$ 
\beq
Z_0=\sum_{n\ge0}{(-1)^n\over n!}\Big[\sum_{l\ge1}{\zeta^{\beta l}\over l}
{1\over\beta}Tr({\bf w}^{\beta l})\Big]^n
\label{rwe}
\eeq
has an interpretation in terms of random walks: the matrix elements $w_{x,x'}$
can be considered as (random) steps in positive time direction between the
lattice sites at $r$ and $r'$. Then the trace $Tr({\bf w}^{\beta l})$
consists of terms which start at a site $r$, go $l\beta$ steps and 
eventually return to $r$. This is a consequence of the fact that ${\bf w}$ 
has non-zero matrix elements only with $t,t+1$. For $l>1$ the walk is sent
back to $t=1$ after $\beta$ steps due to the periodic boundary conditions
defined by the matrix (\ref{matrix}). A realization of random walks is shown 
in Fig.1. The number $l-1$ can be associated with the winding number of the 
corresponding walk.

\subsection{Bose Gas with Intermediate Interaction} 

A weakly-interacting Bose Gas is described by the Bose-Hubbard model
\cite{fisher,sachdev,oosten01}.
This can be derived from the random-walk expansion of Eq. (\ref{rwe}) by
adding an interaction between the random walks. Since the interaction
is local and repulsive, the statistical weight of intersecting walks must
be reduced in comparison with their non-intersecting counterparts. This can
be implemented in the random-walk representation of Eq. (\ref{rwe})
by introducing a phase factor to the tunneling matrix element
\[
w_{x,x'}\to w_{x,x'}e^{2i\varphi_{x}}\equiv v_{x,x'}.
\]
and using for the partition function of the interacting system
\[
Z_{I}=\langle det({\bf 1}-\zeta{\bf v})^{-1}\rangle_\varphi
\]
with the Gaussian distribution of $\varphi_{x}$ with zero mean,
 and 
$\langle\varphi_{x}^2\rangle_\varphi=U$.
The expansion of $Z_{I}$ in powers of $\zeta$ yields terms of the type
\[
\langle\prod_{x'}({\bf v}_{x,x'})^{l_{x'}}\rangle_\varphi
=\prod_{x'}({\bf w}_{x,x'})^le^{-U(\sum_{x'}l_{x'})^2}.
\]
For a singly occupied $x$ (i.e. $\sum_{x'}l_{x'}=1$) the averaging gives a 
rescaling of the fugacity $\zeta\to\zeta e^{-U}$. 
In the case of a single well (i.e., $J=0$ in expression (\ref{matrix}))
the partition function can be easily evaluated and gives with
$\zeta=e^\mu$ ($\mu$ being the chemical potential)
\[
Z_{I}=\Big\langle\Big(1-\zeta^\beta\prod_te^{2i\varphi_{t,t+1}}\Big)^{-1}
\rangle_\varphi
=\sum_{n\ge0}e^{-\beta U(n-\mu/U)n}.
\]
The summation over $n\ge0$, here formally obtained from a geometric series, 
corresponds with the summation over particle
numbers, as one can see by comparing with Eq. (\ref{partition}).
For large $\beta$ only the term with minimal value of $(n-\mu/U)n$
contributes significantly. This is a (discrete) saddle point approximation:
\[
\sum_{n\ge0}e^{-\beta U(n-\mu/U)n}\approx
\max_{n\ge0}e^{-\beta U(n-\mu/U)n}.
\]
The value of $n$ that gives the maximum is also the average number of
particles. To evaluate this quantity one has to determine the groundstate
energy
\[
E_0(\mu/U)=\min_{n\ge0}U(n-\mu/U)n
\]
for a given value of $\mu/U$. The energy $E(\mu/U,n)=U(n-\mu/U)n$
as a function of $\mu/U$ is linear. This implies for the groundstate energy
a piecewise linear function with slope $-n$ for $2n-1\le\mu/U\le2n+1$
(cf. Fig. 2). If one includes tunneling processes, e.g. using an expansion
in small $J$, the phase diagram
would give the lobe structure, discussed for the Bose-Hubbard model
\cite{fisher,sachdev,oosten01}. 

\subsection{Hard-core Bose Gas ($I=1$)}

In the case of hard-core interaction it is not just that random walks have 
to pay a penalty when they intersect but intersections are prohibited 
completely. This is a constraint to the random walks which can be
included explicitly by using the functional integral representation.
Starting from $Z_0$
\[
Z_0={1\over\det({\bf 1}-\zeta {\bf w})}
=\int\exp\Big(-\sum_{x,x'}\Phi_x({\bf 1}-\zeta {\bf w})_{x,x'}
\Phi_{x'}^*\Big)\prod_xd\Phi_xd\Phi^*_x/\pi,
\]
the complex field $\Phi_x$ is replaced by a nilpotent field:
$\Phi_x\to\eta_x^{1}$, $\Phi_x^*\to\eta_x^{2}$, where $(\eta_x^{j})^l=0$
for $l>1$. With the integral over the nilpotent field 
\[
\int_{HC}
\prod_{x\in\L '\subseteq\L}\prod_{\s\in j_{x}}
\eta_x^{\s}
=\cases{
1& if $\L '= \L$, $j_{x}=1,2$\cr
0& otherwise\cr
},
\]
where $\L$ is the full lattice spanned by $x$, the partition function of 
the hard-core Bose gas reads
\[
Z_{HC}=\int_{HC}\exp\Big[\sum_{x,x'}
\eta_x^{1}({\bf 1}+\zeta{\bf w})_{x,x'}\eta_{x'}^{2}\Big].
\]
The expansion of $Z_{HC}$
in powers of $\zeta$ produces the non-intersecting random walks thanks to
the nilpotent condition of the $\eta$'s.

Physical interesting quantities like the density of particles $n$ are
expressed directly in terms of $Z_{HC}$:
\beq
n=-{\zeta\over\beta{\cal N}}{\partial\log Z_{HC}\over\partial\zeta}.
\label{density}
\eeq
The groundstate of a Bose gas can exhibit off-diagonal long-range order
that is characterized by a non-vanishing condensate density $n_0$, obtained
from a correlation function that is related to $Z_{HC}$: 
\[
\lim_{|r|\to\infty}{1\over Z_{HC}}
{1\over\beta}\sum_t\int_{HC}\eta^1_0\eta^2_{(r,t)}\exp\Big[\sum_{x,x'}
\eta_x^{1}({\bf 1}+\zeta{\bf w})_{x,x'}\eta_{x'}^{2}\Big]
=n_0>0.
\]

\section{Fermion Representation of Hard-Core Bosons}

The nilpotent variables $\eta^j_x$ can be replaced by products of
Grassmann variables as $\eta^\s_x\to (-1)^\s\psi^\s_x\bpsi^\s_x$. 
This gives with the usual Grassmann integration \cite{negele} the 
partition function \cite{ziegler94}
\[
Z_{HC}=\int e^{-S_{\rm ferm}}\prod_{x,\s} d\psi_x^{\s}d\bpsi_x^{\s}
\]
with
\[
S_{\rm ferm}=
-\sum_x(\psi_x^{1}\psi_x^{2}+\bpsi_x^{1}\bpsi_x^{2})
-\zeta\sum_{x,x'}w_{x,x'}
\psi_{x}^{1}\bpsi_{x}^{1}\bpsi_{x'}^{2}\psi_{x'}^{2}.
\]
The advantage of this representation is that a quadratic form
of Grassmannians in the exponent can be integrated out, leading to
a determinant. Of course, in the case under consideration $S_{ferm}$
is not a quadratic form. However, the non-quadratic part can be
expressed by a coupling of a quadratic form to an Ising spin $S_{x,x'}=\pm1$:
\[
\exp\Big(\zeta\sum_{x,x'} w_{x,x'}\psi_{x}^{1}\bpsi_{x}^{1}\bpsi_{x'}^{2}
\psi_{x'}^{2}\Big)
\]
\[
=2^{-{\cal N}\beta}\sum_{\{S_{x,x'}=\pm1\}}
\exp\Big(\sum_{x,x'}\sqrt{\zeta w_{x,x'}}
S_{x,x'}(\psi_{x}^{1}\psi_{x'}^{2}+\bpsi_{x}^{1}\bpsi_{x'}^{2})\Big),
\]
where ${\cal N}$ is the number of lattice sites.
Now the Grassmann integration can be performed and gives
\[
Z_{HC}=2^{-{\cal N}\beta}\sum_{\{S_{x,x'}=\pm1\}}\det({\bf 1}+{\bf U})^2
\]
with the matrix elements
\[
U_{x,x'}=\sqrt{\zeta w_{x,x'}}S_{x,x'}.
\]
The density of bosons $n$ is the expectation value
\[
n={1\over2{\cal N}\beta}{\sum_{\{S_{x,x'}\}}\det({\bf 1}+{\bf U})^2
Tr(({\bf 1}+{\bf U})^{-1}{\bf U})\over\sum_{\{S_{x,x'}\}}
\det({\bf 1}+{\bf U})^2}
\]
\[
\equiv{1\over2{\cal N}\beta}\langle Tr(({\bf 1}+{\bf U})^{-1}{\bf U})
\rangle_{S},
\]
with respect to the distribution of Ising spins
\[
P({\bf U})={\det({\bf 1}+{\bf U})^2\over\sum_{\{S_{x,x'}\}}
\det({\bf 1}+{\bf U})^2}.
\]
This representation is a good starting point for a Monte Carlo simulation
of the hard-core Bose gas \cite{hirsch,hettler}.

In one dimension the partition function can be evaluated exactly,
since periodic boundary conditions are satisfied for each random walk
individually (i.e. the winding number is zero) due to the hard-core 
condition. 
Then the determinant is just a sum over all products along possible 
random walks with periodic boundary conditions in $t$:
\[
det({\bf 1}+{\bf U})=1+\sum_{\pi\ne id}\prod_rg_{r,\pi}
\]
with
\[
g_{r,\pi}=\prod_tU_{x,\pi(x)}.
\]
Each $g_{r,\pi}$ contains a product of Ising spins. Now one takes the
square of the determinant and the sum with respect to the 
configurations of the Ising spins. This gives
\[
\sum_{\{S_{x,x'}=\pm1\}}(1+\sum_{\pi\ne id}\prod_rg_{r,\pi})^2
=2^{{\cal N}\beta}+\sum_{\pi\ne id}\sum_{\{S_{x,x'}=\pm1\}}\prod_rg_{r,\pi}^2
=2^{{\cal N}\beta}(1+\sum_{\pi\ne id}\prod_rg_{r,\pi}^2),
\]
since $g_{r,\pi}^2$ does not depend on the Ising spins:
\[
g_{r,\pi}^2=\prod_tU_{x,\pi(x)}^2=\zeta^\beta\prod_t{ w}_{x,\pi(x)}.
\]
Therefore, the partition function can also be written as a determinat
\[
Z_{HC}=\det({\bf 1}+\zeta{\bf w}).
\]
For the occupation of the quantum number $l$ follows
\[
N_l={\zeta^\beta e^{\beta\epsilon_l}\over
1+\zeta^\beta e^{\beta\epsilon_l}}
\sim\cases{
1 & for $\zeta e^{\epsilon_l}>1$ \cr
0 & for $\zeta e^{\epsilon_l}<1$ \cr
}
\]
for $\beta\sim\infty$. Thus a quantum state is either empty or singly occupied,
a consequence of the hard-core interaction. This is also the property of a
non-interacting fermion gas. It reflects the equivalence of the hard-core 
Bose gas and the free fermion gas in one dimension \cite{girardeau}.

\section{Generalization to $N$ States and $N$ Colors}

Generalizations of the hard-core Bose gas have
been proposed that lead to exactly soluble limits: the $N$ state bosons
(NSB) \cite{ziegler94} and the $N$ color bosons (NCB) \cite{ziegler91}.
In the first case a boson can occupy one of $N$ different states
at each lattice site, where these states have the same energy. The hard-core 
interaction applies only to bosons
in the same state. Therefore, bosons can avoid the interaction during
the tunneling process by choosing an unoccupied state. If $N$ is large
the NSB are effectively weakly interacting. Nevertheless, it turns out
that the interaction plays an important role even in the limit $N\to\infty$
\cite{ziegler94}.
The other generalization uses colored bosons: the hard-core interaction
applies only to bosons with the same color. In this situation the interaction
is effectively strong for all values of $N$. In particular, it will be
shown that even in the limit $N\to\infty$ it can destroy the Bose-Einstein
condensate and create a Mott insulator, an effect not possible in the 
limit $N\to\infty$ of the NSB.

\subsection{Effective Field Theory of NSB}

The extension of $S_{ferm}$ to the NSB is given by
\[
S_{NSB}=\sum_x\sum_{\alpha=1}^N
(\psi_x^{1,\alpha}\psi_x^{2,\alpha}+\bpsi_x^{1,\alpha}\bpsi_x^{2,\alpha})
+{\zeta\over N}\sum_{x,x'}w_{x,x'}\sum_{\alpha,\beta=1}^N
\psi_{x}^{1,\alpha}\bpsi_{x}^{1,\alpha}\bpsi_{x'}^{2,\beta}
\psi_{x'}^{2,\beta}.
\]
The Hubbard-Stratonovich transformation
\[
\sum_\alpha\psi_x^{1,\alpha}\psi_x^{2,\alpha} \to\varphi_x,\chi_x
\]
leads to the partition function
\[
Z_{NSB}=\int e^{-N S_{NSB}'}{\cal D}[\vphi,\chi]
\]
with the action
\beq
S_{NSB}'=(\vphi,({\bf 1}+{\bf w})^{-1}\vphi^*)+(\chi,\chi^*)
-\sum_x\log [\z^{-1} +(\vphi_x+i\chi_x)(\vphi_x^*+i\chi_x^*)].
\label{action2}
\eeq
$S_{NSB}'$ does not depend on $N$ and has a $U(1)$ symmetry: 
it is invariant under the transformation
$\varphi_x\to e^{i\eta}\varphi_x$, $\chi_x\to e^{i\eta}\chi_x$, with a 
global phase $\eta$.

\subsection{Effective Field Theory of NCB}

In the case of NCB the action $S_{ferm}$ is extended to
\[
S_{NCB}=\sum_x\sum_{\alpha=1}^N
(\psi_x^{1,\alpha}\psi_x^{2,\alpha}+\bpsi_x^{1,\alpha}\bpsi_x^{2,\alpha})
+{\zeta\over N}\sum_{x,x'}w_{x,x'}\sum_{\alpha,\beta=1}^N
\psi_{x}^{1,\alpha}\bpsi_{x}^{1,\beta}\bpsi_{x'}^{2,\beta}
\psi_{x'}^{2,\alpha}.
\]
The Hubbard-Stratonovich transformation
\[
\sum_\alpha(\psi_x^{1,\alpha}\psi_{x'}^{2,\alpha}
+\bpsi_x^{1,\alpha}\bpsi_{x'}^{2,\alpha})\to u_{x,x'}
\]
leads to the partition function
\beq
Z_{NCB}=\int\exp\Big\{ -N\sum_{x,x'}{(u_{x,x'})^2\over \zeta w_{x,x'}}
+2N\log\det ({\bf 1} + {\bf u})\Big\}
{\cal D}[{\bf u}].
\label{action3}
\eeq
With the choice of ${\bf w}$ given in Eq. (\ref{tunnel})
the action $S$ can also be expressed, after a decomposition of the
matrix field ${\bf u}={\bf u}_\parallel+{\bf u}_\perp$ 
(${\bf u}_\parallel$ is diagonal in space), as
\[
S_{NCB}'={1\over\zeta(1-J)} Tr({\bf u}_\parallel {\bf u}_\parallel^T)
+{6\over\zeta J} Tr({\bf u}_\perp {\bf u}_\perp^T)
-2\log\det({\bf 1}+{\bf u}_\parallel+{\bf u}_\perp)
\]
This is invariant under the orthogonal transformation
\[
{\bf u}\to {\bf O}{\bf u} {\bf O}^T.
\]

\section{$N\to\infty$ Limits}

The new representations of the partition functions depend on $N$ only 
through a prefactor in front of the $N$-independent action.
Therefore, the functional integration can be performed for large values of $N$ 
by the saddle-point (SP) approximation. That means we have to solve the
extremal condition of the action $S$: $\delta S=0$.
Systematic corrections to the SP solution are obtained in a $1/N$ expansion.

\subsection{\bf Results for the NSB}

There are two SP solutions, one is the trivial solution
$\varphi=\chi=0$ which is valid for $\zeta<1$ and a non-trivial
solution with $|\chi|=|\varphi|/2=\sqrt{1-1/\zeta}$ which is valid for 
$\zeta>1$. This implies for total density of bosons at zero temperature
\[
n(\zeta)\sim\cases{
0 & for $\z<1$\cr
1-\z^{-1} & for $\z>1$\cr
}.
\]
Moreover, all bosons are in the condensate in the $N\to\infty$ limit.
However, the $1/N$ corrections indicate a depletion of the 
condensate, as shown in Fig. 3. This phenomenon reflects a partial
destruction of the condensate due to interaction. It can be understood
as a precursor for a new groundstate in the interacting Bose gas,
visible in the NCB as the Mott insulator (s. next section).
The critical point $\zeta=1$ of the Bose-Einstein condensation does not 
depend on the tunneling rate $J$ and agrees with $J=0$ for all $N$. 
It should be noticed that the limit $N\to\infty$ of the NSB leads to the 
Gross-Pitaevskii equation \cite{ziegler02}, at least in a close vicinity of the
critical point (dilute regime).

Quasiparticles are available from $1/N$ corrections. In the condensate
(i.e. $\zeta\ge1$) they have a dispersion on large scales with respect to the
three-dimensional wavevector $k$ as
\[
\epsilon(k)=\sqrt{{J(\z-1)\over3}k^2+(J/6)^2k^4}.
\]
Due to the spontaneously broken $U(1)$ symmetry of the NSB
there is a Goldstone mode $\phi$ for $\zeta>1$ which satisfies the equation
\[
\Big(c^2k^2-\om^2\Big)\phi(k,\omega)=0,
\]
where 
$
c=\sqrt{J(\z-1)/3}
$
is the sound velocity.

\subsection{\bf Results for the NCB}

For the SP equation of the action $S_{NCB}$ one can distinguish
symmetric and symmetry-breaking solutions. Examples of the former are
the trivial solution ${\bf u}_\parallel={\bf u}_\perp=0$ (empty phase) and
${\bf u}_\parallel\ne0$, ${\bf u}_\perp=0$ with identical matrix elements 
\[
u_{x,x'}=\cases{
u_\parallel & for $r'=r$, $t'=t+1$ \cr
0 & otherwise \cr 
}.
\]
This ansatz leads to the mean-field action
\[
{\bar S}_{NCB}={\cal N}\beta\Big[
{u_\parallel^2\over\zeta(1-J)}
-{2\over\beta}\log(1+u_\parallel^\beta)
\Big]
\]
which is minimized by the solution of the SP equation
\[
u_\parallel^2=\zeta (1-J)
{u_\parallel^{\beta}\over1+u_\parallel^\beta}. 
\]
For zero temperature the value of $u_\parallel$ jumps from 0 to 1 at
$\zeta=1/(1-J)$. 
The density of bosons can be calculated with Eq. (\ref{density})
from the mean-field action and gives $n=\Theta(\zeta(1-J)-1)$.
This is the behavior of the system without tunneling. The jump should
be intercepted by an intermediate phase, most likely the Bose-Einstein
condensate. The off-diagonal long-range order of the NCB in the
large-$N$ limit is related to a symmetry-breaking SP solution
with ${\bf u}_\perp\ne 0$. This represents a Bose-Einstein condensate
with a Goldstone mode. It is anticipated that this solution agrees 
qualitatively with the result of the NSB, discussed in the previous section.
The resulting phase diagram, obtained from the $N\to\infty$ limits of the 
NSB and the NCB, is shown in Fig. 4.

\section{Conclusions}

The hard-core Bose gas at zero temperature has been studied in terms of
the soluble $N\to\infty$ limits of NSB and NCB. 
NCB describe three different phases of the hard-core Bose gas,
characterized by a matrix-like order parameter 
${\bf u}={\bf u}_\parallel+{\bf u}_\perp$. There is an {\it empty phase} 
(${\bf u}_\parallel={\bf u}_\perp=0$), a {\it Bose-Einstein condensate}
(${\bf u}_\parallel=0, {\bf u}_\perp\ne0$),
and a {\it Mott insulator} (${\bf u}_\parallel\ne0, {\bf u}_\perp=0$).
In contrast to the NCB  the complex scalar order
parameter of the NSB describes only
an empty phase and a Bose-Einstein condensate but not the Mott insulator.
Thus, the NSB has only one symmetric SP solution (empty phase), 
whereas the NCB has at least two (empty phase and Mott insulator).
However, the depletion of the condensate in $o(1/N)$ of the NSB indicates 
the partial destruction of the latter. The main result of this study is the 
effective action of the NCB of the matrix field ${\bf u}$
\[
S_{NCB}'={1\over\zeta(1-J)} Tr({\bf u}_\parallel {\bf u}_\parallel^T)
+{6\over\zeta J} Tr({\bf u}_\perp {\bf u}_\perp^T)
-2\log\det({\bf 1}+{\bf u}_\parallel+{\bf u}_\perp),
\]
where its minima represent different phases of the strongly interacting Bose 
gas.

\vskip0.5cm
\no
{\bf Acknowledgement}

\no
The author is grateful to Ch. Moseley for useful discussions.

\newpage

\begin{figure} 
\begin{center}
\includegraphics{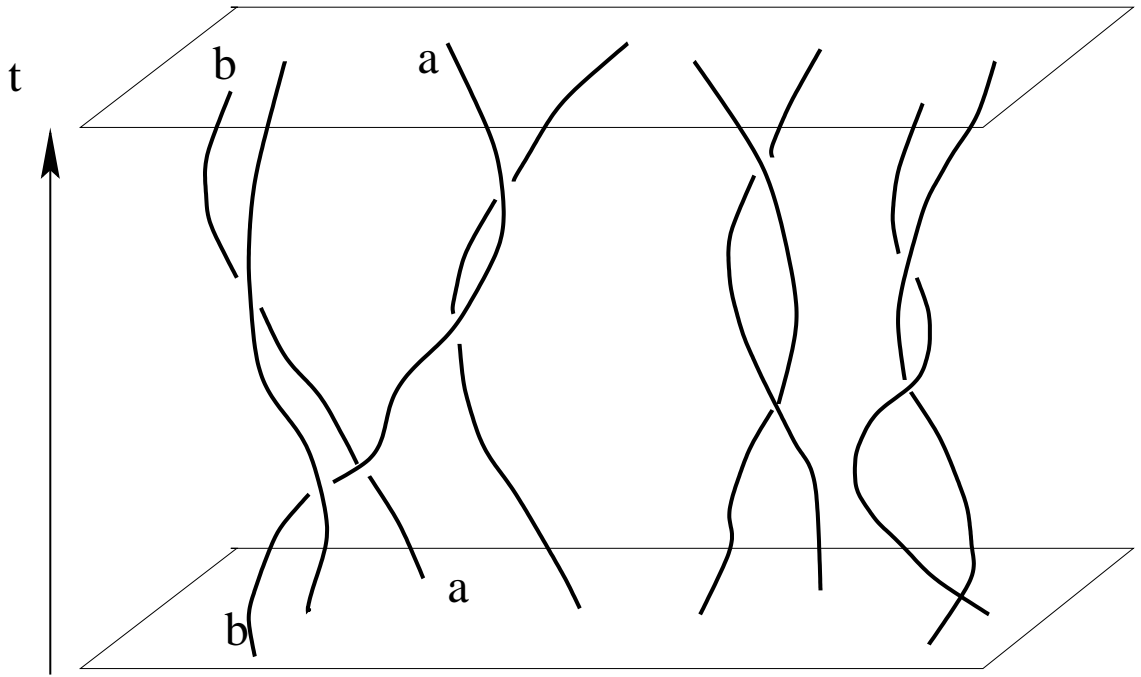}
\end{center} 
\caption{Contribution of the random-walk expansion to the partition
function. Sites $a$ and $b$ are connected by the periodic boundary 
conditions, a consequence of the fact that the partition function
is a trace (cf. text).
The hard-core interaction does not allow intersecting random walks.}
\end{figure}

\begin{figure} 
\begin{center}
\includegraphics{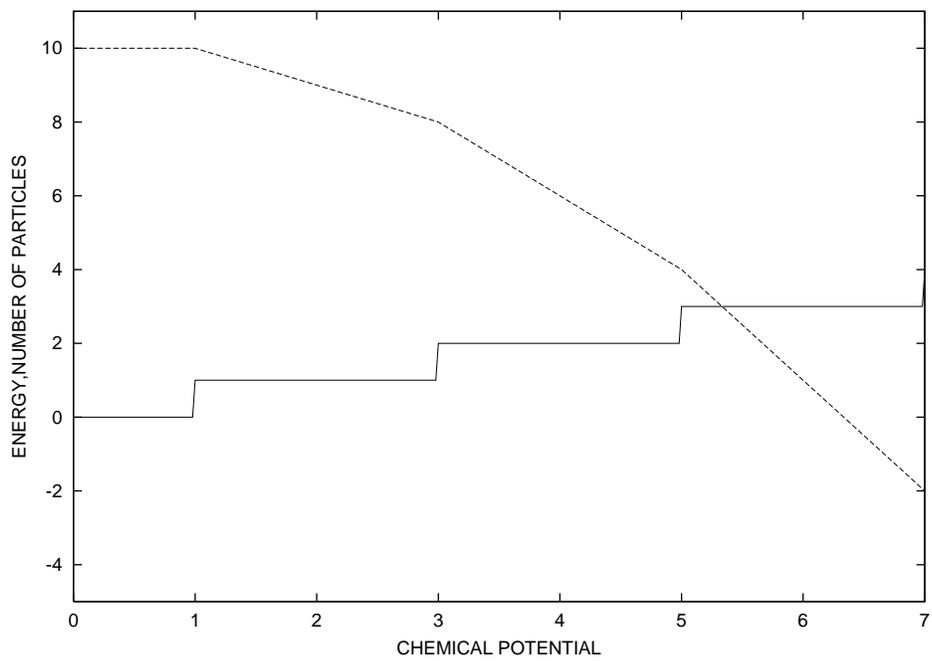}
\end{center} 
\caption{Groundstate energy and step-like density of particles of an 
interacting Bose gas in a single well as a function of $\mu/U$.
}
\end{figure}

\begin{figure} 
\begin{center}
\includegraphics{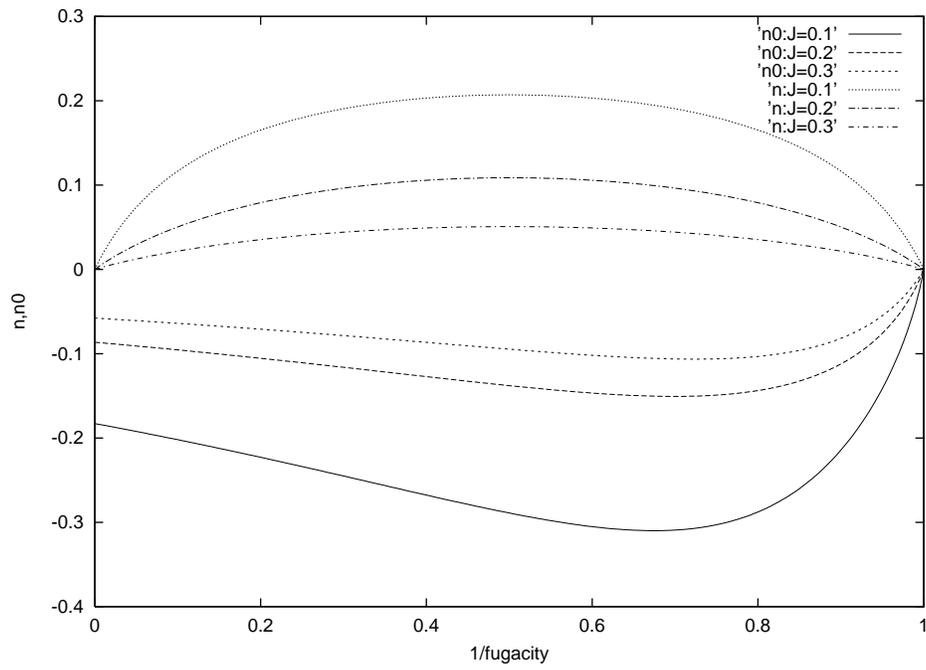}
\end{center} 
\caption{$1/N$ corrections of $N$ state bosons: 
enhancement of the total density $n$ 
(upper three curves) and depletion of the condensate density $n_0$ (lower 
three curves) \cite{moseley02}.}
\end{figure}

\begin{figure} 
\begin{center}
\includegraphics{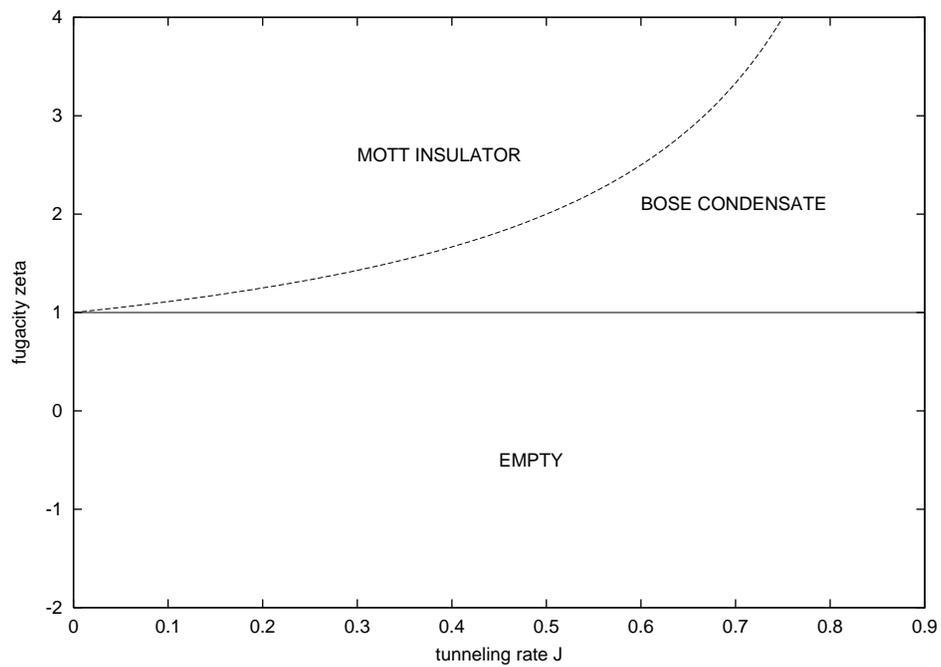}
\end{center} 
\caption{$T=0$--phase diagram obtained from $N\to\infty$ limits. The phase
boundary to the Mott insulator was calculated with $N$ color bosons, the 
phase boundary to the empty phase with $N$ state bosons.}
\end{figure}

\end{document}